\documentclass[showpacs,twocolumn,aps,prb,a4paper]{revtex4-1}

\usepackage{color,epsfig,rotating,graphicx,subfigure}
\usepackage{amsmath,amssymb,amscd,afterpage,ulem}

\usepackage[latin1]{inputenc}
\usepackage[english]{babel}

\usepackage{dsfont}
\usepackage{textcomp}

\renewcommand{\Re}{{\rm Re}}
\renewcommand{\Im}{{\rm Im}}

\newcommand{\ri}{{\rm i}}
\newcommand{\re}{{\rm e}}
\newcommand{\rd}{{\rm d}}

\newcommand{\rs}{{\rm s}}
\newcommand{\rp}{{\rm p}}
\newcommand{\kb}{k_{\rm B}}

\newcommand{\rI}{{\rm I}}

\newcommand{\ra}{{\rm a}}
\newcommand{\rb}{{\rm b}}

\begin{document}

\title{Anisotropy enhancement of the Casimir-Polder force between a nanoparticle and graphene}
\author{S.-A. Biehs$^*$}

\affiliation{Institut f\"{u}r Physik, Carl von Ossietzky Universit\"{a}t, D-26111 Oldenburg, Germany\\
$^*$Corresponding author: s.age.biehs@uni-oldenburg.de}

\author{G. S. Agarwal}

\affiliation{Department of Physics, Oklahoma State University, Stillwater, Oklahoma 74078, USA}

\date{\today}

\begin{abstract}
We derive the analytical expressions for the thermal Casimir-Polder energy and force between a spheroidal 
nanoparticle above a semi-infinite material and a graphene covered interface. We analyze in detail
the Casimir-Polder force between a gold nanoparticle and a single sheet of pristine graphene focusing on 
the impact of anisotropy. We show that the effect of anisotropy, i.e.\ the shape and orientation of the
spheroidal nanoparticle, has a much larger influence on the force than the tunability of graphene. The effect
of tuning and anisotropy both add up such that we observe a force between the particle and the sheet of
graphene which is between 20-50\% of that between the same particle and an ideal metal plate.
Hence, the observed force is much larger than the results found for the Casimir force between a metal halfspace
and a layer of graphene.
\end{abstract}


\maketitle
\newpage

%
%
%

\section{Introduction}

The interaction of atoms and/or nanoparticles with an interface or a cavity is a research
topic which has attracted a lot of attention in the past and which is still a vital field
of research. Such interactions include for example the change of the radiative life time or Purcell 
effect~\cite{Novotny2012} close to an interface or in a cavity, the energy transfer between a 
nanoparticle and a surface~\cite{Doro1998,MuletEtAl2001} or between two anisotropic nanoparticles~\cite{IncardoneEtAl2014,Nikbakht}, the radiative cooling rate of nanoparticles
in close vicinity to a plasmonic system~\cite{TschikinEtAl2012}, the F\"{o}rster resonance energy transfer in the 
presence of an interface~\cite{GerstenNitzan1984,HuaEtAl1985}, the Spin-Hall effect close to plasmonic systems~\cite{AgarwalBiehs2013} as well 
as the Casimir-Polder (CP) force~\cite{MilonniBook,EmigEtAl2009,LevinEtAl2010,EberleinZietal2011}. 

In particular, the possibility to use materials with special properties as graphene, for 
instance, has renewed the interest in such studies. So it was shown that graphene
allows for controlling the spontaneous emission or local density of states~\cite{KoppensEtAl2011,MessinaEtAl2013} 
and can enhance the radiative heat transfer between two materials~\cite{Svetovoy2011,IlicEtAl2012, MessinaEtAl2013b,LimEtAl2013} as well as 
the F\"orster energy transfer between two atoms in close vicinity of a sheet of graphene~\cite{AgarwalBiehs2013,BiehsAgarwal2013}. Due
to the possibility of changing the electron density of graphene by gating or doping~\cite{CastroNetoEtAl2009} the 
magnitude of these effects can be controlled to a certain extent. Regarding the Casimir force between two or several sheets of 
graphene~\cite{GomezSantos2009,DrosdoffWoods2010,Sernelius2011}, a sheet of graphene and a 
metal~\cite{BordagEtAl2009,Sernelius2011,FialkovskyEtAl2011,Sernelius2012} it turns out that it is 
on the order of some percent of the Casimir force between two perfect metals~\cite{BordagEtAl2009}. Similar
results were found for the CP force between a rubidium atom and a graphene layer in Ref.~\cite{RibeiroScheel} reporting
a value for the CP force of about five percent of that for an ideal metal.
In addition, it could be shown that the effect of gating or doping has only small influence on the Casimir force for 
gapless graphene~\cite{Sernelius2011,Sernelius2012,SvetovoyEtAl2011}. Finally, it could  be shown that by applying 
external magnetic fields the Casimir force between two graphene layers can be completely 
suppressed or even made repulsive due to the quantum Hall effect~\cite{WangKongTse2012} which might 
be useful in the search for Yukawa-like corrections to Newtonian gravity~\cite{BezerraEtAl2010}.

\begin{figure}[Hhbt]
  \epsfig{file = 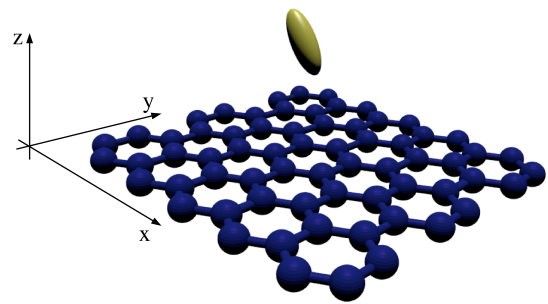, width = 0.45\textwidth}
  \caption{(Color online) Sketch of a spheroidal gold nanoparticle above a sheet of graphene.\label{Sketch}}
\end{figure}

The values obtained for the Casimir force between two or several sheets of graphene or graphene and a metal
which are only some percent of that for perfect metals seem to be rather small values. However, for a 
single-atomic layer such as graphene this magnitude of the Casimir force is astonishingly large and can be 
explained by the unique properties of graphene. But as we show in the following even much higher values
can be obtained for the CP force between a spheroidal nanoparticle and a sheet of graphene
as sketched in Fig.~\ref{Sketch}. It turns out that in the high-temperature limit, i.e.\ for distances 
larger than the thermal wavelength $\lambda_{\rm th} = \hbar c/ \kb T$ ($\lambda_{\rm th}$ is about $7.6\,\mu {\rm m}$ at $T = 300\,{\rm K}$), 
the sheet of graphene essentially acts like a perfect metal regardless of the shape and orientation of the 
particle as was also previously found for atoms above a sheet of 
graphene~\cite{KlimchitskayaMostepanenko2014}. On the other hand, for distances smaller than the thermal 
wavelength such a large CP force persists. Due to the effect of tuning in graphene and the particle's anisotropy 
it can still be as large as 20-50\% of that between a nanoparticle and a perfect metal. Compared to the results 
previously found for the Casimir force these values are extremely large. 

The paper is organized as follows: In Sec.~II we derive the general expressions for the 
CP energy and force between an anisotropic nanoparticle and a halfspace or a graphene-covered
halfspace. In Sec.~III we introduce the models describing the material properties of gold
and graphene which are used in the numerical simulations. Then in Sec.~IV and V we discuss
numerical results for the CP force between isotropic and spheroidal nanoparticles and
a sheet of graphene. Finally, we summarize our results in Sec.~VI.

%
%
%

\section{Casimir-Polder force}

In this section we derive the expression for the CP energy as well as
for the CP potential of an in general anisotropic nanoparticle and an
interface.

\subsection{Interaction energy}

Following the procedure in Ref.~\cite{Novotny2012}, the interaction energy between a small particle and
a surface induced by fluctuations can be written as
\begin{equation}
  H_\rI = - \frac{1}{2} \langle \mathbf{p}^{\rm ind} \cdot \mathbf{E}^{\rm fl} \rangle - \frac{1}{2} \langle \mathbf{p}^{\rm fl}\cdot\mathbf{E}^{\rm ind} \rangle,
\end{equation}
where the induced dipole moment and field in Fourier space are given by
\begin{align}
  \mathbf{p}^{\rm ind}(\omega) &= \epsilon_0 \uuline{\alpha}(\omega)\cdot\mathbf{E}^{\rm fl}(\omega), \\
  \mathbf{E}^{\rm ind}(\omega) &= \omega^2 \mu_0 \mathds{G}(\omega) \cdot \mathbf{p}^{\rm fl}(\omega).
\end{align}
Then we obtain (using Einstein's convention)
\begin{equation}
\begin{split}
  H_\rI &= - \Re \int_0^\infty \!\frac{\rd \omega}{2 \pi} \epsilon_0  
          \alpha_{ij}(\omega) \langle E^{\rm fl}_j (\omega) {E^{\rm fl}_i}^* (\omega)\rangle\\
        &\quad - \Re\int_0^\infty \!\frac{\rd \omega}{2 \pi} \omega^2 \mu_0 
          \mathds{G}_{ij}^* (\omega) \langle p^{\rm fl}_i (\omega) {p^{\rm fl}_j}^* (\omega)\rangle 
\end{split}
\end{equation}
taking into account that $\langle p_i^{\rm fl} p_j^{\rm fl}\rangle = \langle {p_i^{\rm fl}}^* {p_j^{\rm fl}}^*\rangle = 0$
and $\langle E_i^{\rm fl} E_j^{\rm fl}\rangle = \langle {E_i^{\rm fl}}^* {E_j^{\rm fl}}^*\rangle = 0$. Furthermore,
we have already used the fact that the correlation functions are delta correlated with respect to the frequency due to the
stationarity of the equilibrium situation. Using the fluctuation-dissipation theorem of second and first kind
\begin{align}
  \langle p^{\rm fl}_i (\omega) {p^{\rm fl}_j}^* (\omega)\rangle &= \Theta(\omega,T) \frac{2 \epsilon_0}{\omega} \Im [\alpha_{ij}(\omega)], \\
  \langle E^{\rm fl}_j (\omega) {E^{\rm fl}_i}^* (\omega)\rangle &= \Theta(\omega,T)  2 \mu_0 \omega \Im [\mathds{G}_{ij}(\omega)]
\end{align}
where
\begin{equation}
  \Theta(\omega, T) = \frac{\hbar \omega}{2} + \frac{\hbar \omega}{\re^{\hbar \omega \beta} - 1}
\end{equation}
we arrive at
\begin{equation}
   H_\rI = - \Im \int_0^\infty \!\!\rd \omega \, \frac{\omega}{\pi c^2} \Theta(\omega,T) \mathds{G}_{ij} \alpha_{ij}.
\end{equation}
Finally, we perform a Wick rotation ($\omega \rightarrow \ri \xi$) for zero and nonzero temperatures assuming that the polarizability and 
the Green's function do not have any poles or branchpoints inside the first quadrant. Then we obtain
\begin{align}
  H_\rI^{T = 0} &= \int_0^\infty \!\! \rd \xi\, \frac{\hbar \xi^2}{\pi c^2} \alpha_{ij}(\ri \xi) \mathds{G}_{ij}(\ri \xi), \\
  H_\rI^{T \neq 0} &= {\sum_{n=0}^\infty}' \frac{2 \kb T}{c^2} \xi^2_n \alpha_{ij}(\ri \xi_n) \mathds{G}_{ij}(\ri \xi_n)
\end{align}
introducing the Matsubara frequencies $\xi_n = 2 \pi n \frac{\kb T}{\hbar}$. The prime at the sum sign symbolizes that the
term for $n = 0$ has to be multiplied by $1/2$. Note, that any magnetic response even that of eddy currents has been 
neglected~\cite{Tomchuk2006} which can be important for thermal heat transfer~\cite{ChapuisEtAl2008,HuthEtAl2010}. 

\subsection{Green's function}

Now, we assume that the nanoparticle is in front of a planar medium at distance $d$. 
For determining the CP force only the scattered part of the Green's function is needed, since we are interested in 
the energy difference $H_\rI (d) - H_\rI (d \rightarrow \infty)$. The scattering part of the Green's function 
close to a planar interface is~ \cite{Novotny2012} (for $\mathbf{x} = \mathbf{x}'$ and $z = z' = d$)
\begin{equation}
  \mathds{G}^{(\rm sc)} = \int_0^\infty \!\! \frac{\rd \kappa}{2 \pi} \kappa \frac{\ri \re^{\ri \gamma_0 2 d}}{2 \gamma_0} \bigl[ r_\rs \mathds{S} + r_\rp \mathds{P} \bigr]
\end{equation}
with
\begin{equation}
  \mathds{S} = \begin{pmatrix} \frac{1}{2} & 0 & 0 \\ 0 & \frac{1}{2} & 0 \\ 0 & 0 & 0 \end{pmatrix}
\end{equation}
and
\begin{equation}
  \mathds{P} = \begin{pmatrix} - \frac{1}{2} \frac{\gamma_0^2}{k_0^2} & 0 & 0  \\
                               0 & - \frac{1}{2} \frac{\gamma_0^2}{k_0^2}  & 0  \\
                               0 & 0 & \frac{1}{2} \frac{\kappa^2}{k_0^2}
\end{pmatrix}.
\end{equation}
Here $r_\rs$ and $r_\rp$ are the usual Fresnel reflection coefficients for s- and p-polarized light
\begin{equation}
  r_\rs = \frac{\gamma_0 - \gamma}{\gamma_0 + \gamma} \qquad \text{and} \qquad r_\rp = \frac{\gamma_0 \epsilon - \gamma}{\gamma_0 \epsilon + \gamma}
\end{equation}
introducing the wave vector components along the surface normal inside vacuum ($\gamma_0$) and
inside a medium ($\gamma$) having the permittivity $\epsilon$  which are explicitely given by
\begin{equation}
  \gamma_0 = \sqrt{\frac{\omega^2}{c^2} - \kappa^2} \qquad \text{and} \qquad \gamma = \sqrt{\frac{\omega^2}{c^2} \epsilon(\omega) - \kappa^2}.
\end{equation}

\subsection{Casimir-Polder energy and force}

Using the scattered Green's function we obtain for the CP energy performing a Wick rotation ($\omega \rightarrow \ri \xi$)
\begin{align}
  H_{\rm CP}^{T = 0} &= H_\rI^{T = 0} (d) - H_\rI^{T = 0} (d \rightarrow \infty) \nonumber \\ 
                     &= \int_0^\infty \!\! \rd \xi\, \frac{\hbar}{4 \pi^2} f(\xi), \label{Eq:HT0} \\
  H_{\rm CP}^{T \neq 0} &=  H_\rI^{T \neq 0} (d) - H_\rI^{T \neq 0} (d \rightarrow \infty) \nonumber \\ 
                        &={\sum_{n=0}^\infty}' \frac{\kb T}{2 \pi} f(\xi_n)  \label{Eq:HTnot0}
\end{align}
with 
\begin{equation}
\begin{split}
  f(\xi) &=  \int_0^\infty \!\!\rd \kappa \kappa \, \frac{\re^{- \tilde{\gamma}_0 2 d}}{2 \tilde{\gamma}_0} \biggl[ \frac{\xi^2}{c^2} r_\rs(\ri \xi) \frac{\alpha_x + \alpha_y}{2} \\
         &\quad - r_\rp(\ri \xi) \biggl( \frac{\alpha_x + \alpha_y}{2} \tilde{\gamma}_0^2 + \kappa^2 \alpha_z \biggr) \biggr]
\end{split}
\end{equation}
The resulting force is given by
\begin{align}
  F_{\rm CP}^{T = 0} &= - \frac{\partial}{\partial d} H_{\rm CP}^{T = 0} =  \int_0^\infty \!\! \rd \xi\, \frac{\hbar}{4 \pi^2} \tilde{f}(\xi),\\ 
  F_{\rm CP}^{T \neq 0} &= - \frac{\partial}{\partial d} H_{\rm CP}^{T \neq 0} =   {\sum_{n=0}^\infty}' \frac{\kb T}{2 \pi} \tilde{f}(\xi_n), \label{Eq:FCP}
\end{align}
where
\begin{equation}
\begin{split}
  \tilde{f}(\xi) &= \int_0^\infty \!\!\rd \kappa \kappa \, \re^{- \tilde{\gamma}_0 2 d} \biggl[ \frac{\xi^2}{c^2} r_\rs(\ri \xi) \frac{\alpha_x + \alpha_y}{2} \\
          &\qquad - r_\rp(\ri \xi) \biggl( \frac{\alpha_x + \alpha_y}{2} \tilde{\gamma}_0^2 + \kappa^2 \alpha_z \biggr) \biggr]
\end{split}
\end{equation}
with $\tilde{\gamma}_0 = \sqrt{\frac{\xi^2}{c^2} + \kappa^2}$.

\subsection{Isotropic nanoparticle}

For an isotropic nanoparticle we have
\begin{equation}
  \uuline{\alpha}(\ri \xi) = \alpha(\ri \xi) \mathds{1},
\end{equation}
where $\mathds{1}$ is the unit matrix.
For a spherical nanoparticle the polarizability is determined by the 
Mie coefficients. For nanoparticles having a radius smaller than the skin depth
the polarizability can be approximated by the Clausius-Mosotti like
expression
\begin{equation}
  \alpha (\ri \xi) = 4 \pi R^3 \frac{\epsilon(\ri \xi) - 1}{\epsilon(\ri \xi) + 2}.
\label{Eq:ClausiusMosotti}
\end{equation}
The functions $f$ and $\tilde{f}$ reduce in this case to
\begin{equation}
  f(\xi) =  \alpha(\ri \xi) \int_0^\infty \!\!\rd \kappa \kappa \, \frac{\re^{- \tilde{\gamma}_0 2 d}}{2 \tilde{\gamma}_0} \biggl[ \frac{\xi^2}{c^2} r_\rs(\ri \xi) 
            - r_\rp(\ri \xi)  \biggl( \frac{\xi^2}{c^2} + 2 \kappa^2 \biggr) \biggr]
\end{equation}
and
\begin{equation}
   \tilde{f}(\xi) =  \alpha(\ri \xi) \int_0^\infty \!\!\rd \kappa \kappa \, \re^{- \tilde{\gamma}_0 2 d} \biggl[ \frac{\xi^2}{c^2} r_\rs(\ri \xi)  - r_\rp(\ri \xi) \biggl( \frac{\xi^2}{c^2} + 2 \kappa^2 \biggr) \biggr].
\end{equation}


If the medium is an ideal metal, we have $r_\rs = -1$ and $r_\rp = 1$ so that
\begin{equation}
  f_{\rm IM}(\xi) =  - \alpha(\ri \xi) \int_0^\infty \!\!\rd \kappa \kappa \, \frac{\re^{- \tilde{\gamma}_0 2 d}}{2 \tilde{\gamma}_0} \biggl[ 2 \frac{\xi^2}{c^2} + 2 \kappa^2 \biggr]
  \label{Eq:FHIM}
\end{equation}
and
\begin{equation}
   \tilde{f}_{\rm IM}(\xi) =  - \alpha(\ri \xi) \int_0^\infty \!\!\rd \kappa \kappa \, \re^{- \tilde{\gamma}_0 2 d} \biggl[ 2 \frac{\xi^2}{c^2} + 2 \kappa^2 \biggr].
\label{Eq:FCPIM}
\end{equation}
Both equations can be integrated giving
\begin{align}
  f_{\rm IM}(\xi) &= - \alpha(\ri \xi) \frac{\re^{-\frac{\xi}{\omega_{\rm c}}} }{(2d)^3} \biggl[ \biggl(\frac{\xi}{\omega_{\rm c}}\biggr)^2 + 2 \frac{\xi}{\omega_{\rm c}} + 2 \biggr], \\
 \tilde{f}_{\rm IM}(\xi) &=  - \alpha(\ri \xi) \frac{2\re^{-\frac{\xi}{\omega_{\rm c}}}}{(2d)^4}  \biggl[ \biggl(\frac{\xi}{\omega_{\rm c}}\biggr)^3 + 3 \biggl(\frac{\xi}{\omega_{\rm c}}\biggr)^2 + 6 \frac{\xi}{\omega_{\rm c}} + 6 \biggr],
\end{align}
where we have introduced the characteristic frequency $\omega_{\rm c} \equiv c/(2d)$.
Inserting these functions into the expressions for the CP
potential or force gives the corresponding results for a spherical nanoparticle 
above a perfect metal~\cite{BabbEtAl2004,ScheelBuhmann2008}.

\subsection{Spheroidal nanoparticle}

Now, let us assume that we have a spheroidal nanoparticle with radii 
$R_z \equiv R_\ra$ and $R_x = R_y \equiv R_\rb$, i.e.\ the rotational axis is along the z axis. The nonzero components of the 
polarizability tensor are then given by the diagonal elements ($i = x,y,z$)
\begin{equation}
  \alpha_i (\ri \xi) = \frac{4 \pi}{3} R_\ra R_\rb^2 \frac{\epsilon(\ri \xi) - 1}{1 + [\epsilon(\ri \xi) - 1] L_i }
  \label{Eq:PolEllipsoidal}
\end{equation}  
where the depolarization factors $L_i$ are for a spheroidal nanoparticle given by the analytical 
expressions~\cite{Landau}
\begin{align}
  L_x &= L_y = \frac{1}{2} (1 - L_z) \\
  L_z &= \begin{cases}
           \frac{1 - e^2}{e^2} \biggl[\frac{1}{2 e}\ln\biggl(\frac{1 + e}{1 - e} \biggr)  - 1 \biggr], & R_\ra > R_\rb \\
           \frac{1 + e^2}{e^2} \biggl( 1 - \frac{\arctan(e)}{e}\biggr), & R_\ra < R_\rb  
         \end{cases}.
\end{align}
Note that the expressions for $L_z$ are different for oblate ($R_a < R_b$) and prolate ($R_a > R_b$) nanoparticles
as well as the expressions for the eccentricity 
\begin{equation}
  e^2 \equiv \begin{cases}
             1 - \frac{R_\rb^2}{R_\ra^2}  , & R_\ra > R_\rb \\
             \frac{R_\rb^2}{R_\ra^2} - 1 , & R_\ra < R_\rb
           \end{cases}.
\end{equation}
Although it would be an easy task to consider all orientations of the spheroidal nanoparticle with respect to a surface, for convenience and clarity 
we will in the following focus on the CP force between spheroidal nanoparticles above a surface where the rotational axes of the 
particles are normal or parallel to the interface. 

For the case that the surface is replaced by an ideal metal, the above expressions for $f$ and $\tilde{f}$ which enter
in the CP energy and force formulas reduce to
\begin{align}
f_{\rm IM}(\xi) &=  - \int_0^\infty \!\!\rd \kappa \kappa \, \frac{\re^{- \tilde{\gamma}_0 2 d}}{2 \tilde{\gamma}_0} \biggl[ \biggl( 2 \frac{\xi^2}{c^2} + \kappa^2\biggr) \frac{\alpha_x + \alpha_y}{2} + \kappa^2 \alpha_z \biggr], \\
\tilde{f}_{\rm IM}(\xi) &= - \int_0^\infty \!\!\rd \kappa \kappa \, \re^{- \tilde{\gamma}_0 2 d} \biggl[ \biggl(2 \frac{\xi^2}{c^2} + \kappa^2\biggr) \frac{\alpha_x + \alpha_y}{2} + \kappa^2 \alpha_z \biggr].
\end{align}
Note that depending on the orientation of the rotational axis of the spheroidal nanoparticle with respect to the the
interface these expressions can be further simplified. 

%
%
%

\section{Material properties}

In the following we use the above derived expressions to evaluate the CP force between a spherical
nanoparticle made of gold and a sheet of graphene. We will compare our results to the CP
force for the case that the sheet of graphene is replaced by a gold halfspace. Before presenting the numerical
results we introduce the models describing the material properties of the particle and the interface.

\subsection{Gold}

For gold we use for convenience the Drude model given by~\cite{AshcroftBook} 
\begin{equation}
  \epsilon_{\rm Au} (\ri \xi) = 1 + \frac{\omega_\rp^2}{\xi(\xi + \gamma)}
\end{equation}
with $\omega_\rp = 1.4\cdot10^{16}\,{\rm rad} / {\rm s}$ and $\gamma = 3\cdot10^{13}\,{\rm rad} / {\rm s}$.

\subsection{Graphene}

\begin{figure}[Hhbt]
  \epsfig{file = 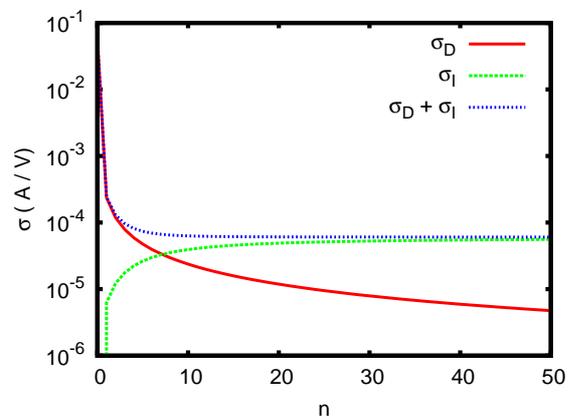, width = 0.45\textwidth}
  \caption{(Color online) Plot of $\sigma = \sigma_D + \sigma_I$, $\sigma_D$, and $\sigma_I$ as a function of the Matsubara terms.\label{Fig:SigmaGraphene}}
\end{figure}

The material properties of pristine graphene in the local limit (for the considered distance regime nonlocal effects can be safely neglected as will be shown in Fig.~\ref{Fig:SpheroidalAuIMDist}) for real 
frequencies are for the Drude-like term $\sigma_D$ and the interband contribution $\sigma_I$~\cite{FalkovskyVarlamov2007,Falkovsky} 
\begin{align}
  \sigma_D &= \frac{\ri}{\omega + \ri/\tau} \frac{2 e^2 \kb T}{\pi \hbar^2} \log\biggl[ 2 \cosh\biggl( \frac{E_F}{2 \kb T} \biggr) \biggr], \\
  \sigma_I &= \frac{e^2 \omega}{\ri \pi \hbar} \biggl[ - \int_0^\infty \rd \varepsilon \, \frac{f_0(-\varepsilon) - f_0(\varepsilon)}{(\omega + \ri \delta)^2 - 4 \varepsilon^2} \biggr],
\end{align}
where
\begin{equation}
  f_0 = \frac{1}{\re^{(\hbar \varepsilon - E_F) \beta} + 1}.
\end{equation}
The Fermi level $E_F$ equals zero for pristine graphene, but it can be changed by controlling the density and type of carriers in graphene by electrical gating, chemical doping or 
substitional doping~\cite{CastroNetoEtAl2009}. Values up to $1\,{\rm eV}$ were reported(see~\cite{GrigorenkoEtAl2012} and references therein).
The resulting values of $\sigma = \sigma_I + \sigma_D$ for $E_F = 0.5\,{\rm eV}$ and $T = 300\,{\rm K}$ are plotted in Fig.~\ref{Fig:SigmaGraphene} as a 
function of the Matsubara terms counted by $n$ (the corresponding Matsubara frequency is $\xi_n = 2 \pi n \frac{\kb T}{\hbar}$) using a moderate damping of $\tau = 10^{-12}\,{\rm rad/s}$~\cite{KoppensEtAl2011}. It can be seen that for small frequencies ($n < 7$) the intraband contribution dominates whereas for large frequencies ($n > 7$) the interband contribution dominates
the conductivitiy of graphene and converges to $e^2/4 \hbar$~\cite{FalkovskyVarlamov2007,StauberEtAl2008}. By changing the Fermi energy $E_F$ this crossover between 
the inter- and intraband contribution can be shifted towards lower or larger Matsubara frequencies.

The reflection coefficients for graphene are different from the expression for a halfspace. On the imaginary axis they are given by~\cite{KoppensEtAl2011}
\begin{equation}
  r_\rs = \frac{\tilde{\gamma}_0 - \tilde{\gamma} - \mu_0 \sigma \xi}{\tilde{\gamma}_0 + \tilde{\gamma} + \mu_0 \sigma \xi} \quad\text{and}\quad
  r_\rp = \frac{\tilde{\gamma}_0 \epsilon - \tilde{\gamma} + \frac{\sigma \tilde{\gamma} \tilde{\gamma}_0}{ \varepsilon_0 \xi}}{\tilde{\gamma}_0 \epsilon + \tilde{\gamma} + \frac{\sigma \tilde{\gamma} \tilde{\gamma}_0}{ \varepsilon_0 \xi}}, 
\end{equation}
where $\epsilon$ is in this case the permittivity of the substrate. For a suspended sheet of graphene $\epsilon = 1$ and $\tilde{\gamma} = \tilde{\gamma}_0$.

%
%
%

\section{Numerical results and discussion - isotropic nanoparticle}

\subsection{Gold}

\begin{figure}[Hhbt]
  \epsfig{file = 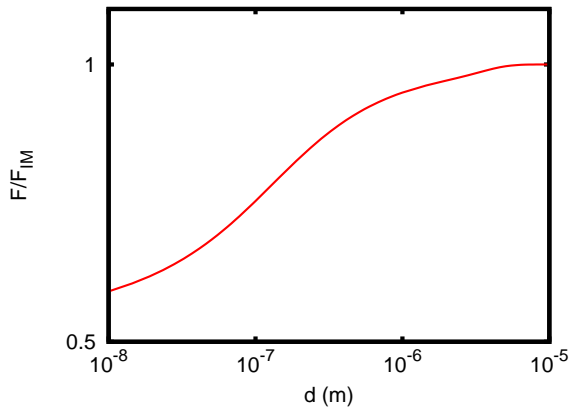, width = 0.45\textwidth}
  \caption{(Color online) Plots of the CP force of a Au nanoparticle above a Au halfspace as a function of distance $d$.
           The force is normalized to the CP force of a gold nanoparticle above an ideal metal.
           Here and in the following we set $T = 300\,{\rm K}$.\label{Fig:AuSiCIM}}
\end{figure}

In Fig.~\ref{Fig:AuSiCIM} we first show our results for a spherical gold nanoparticle above a gold halfspace.
The resulting values of the CP force are normalized to the case, where the gold halfspace is
replaced by an ideal metal, i.e.\ we use expression (\ref{Eq:FCP}) with (\ref{Eq:FCPIM}), while the properties
of the nanoparticles remain unchanged. Due to this normalization procedure the results are independent of the radius 
of the nanoparticles, but it has to be kept in mind that the presented results are only meaningful for distances
$d$ larger than the radius of the nanoparticle. In addition, by using the expressions (\ref{Eq:ClausiusMosotti}) and (\ref{Eq:PolEllipsoidal})
for the polarizability it is assumed for convenience that the nanoparticle is smaller than the skin depth.
This sets another constraint on the validity of the shown results. With the here used Drude parameters we find a 
minimal skin depth $\delta_{\rm s}$ of about $21\, {\rm nm}$. Therefore, the numerical results presented here
are strictly valid for nanoparticles smaller then $21\,{\rm nm}$, only.

From the numerical results it can be seen that at large distances the CP force for the gold halfspace and the ideal 
metal are the same. For smaller distances the CP force drops with respect to the ideal metal case. Hence, by using 
a gold halfspace instead of an ideal metal, the force is reduced by about 25\% for distances around $100\,{\rm nm}$.     

\subsection{Suspended graphene}

\begin{figure}[Hhbt]
  \epsfig{file = 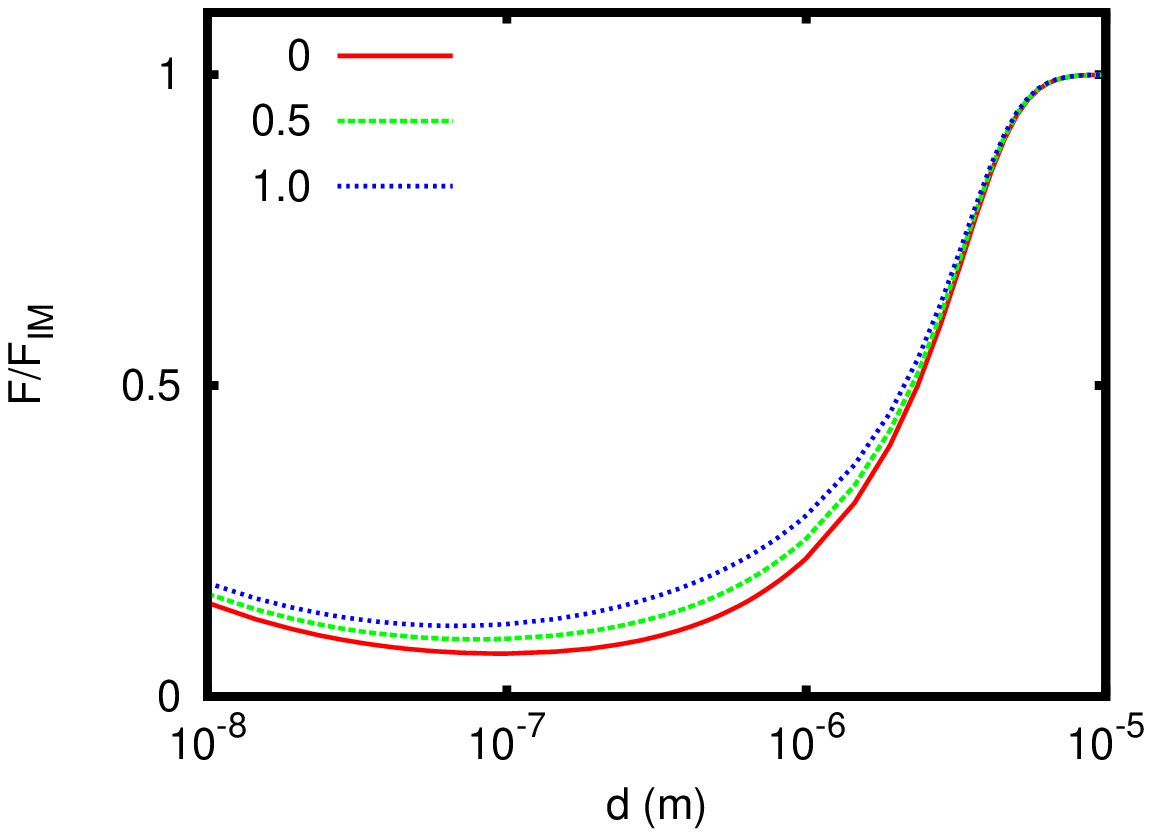, width = 0.45\textwidth}
  \epsfig{file = 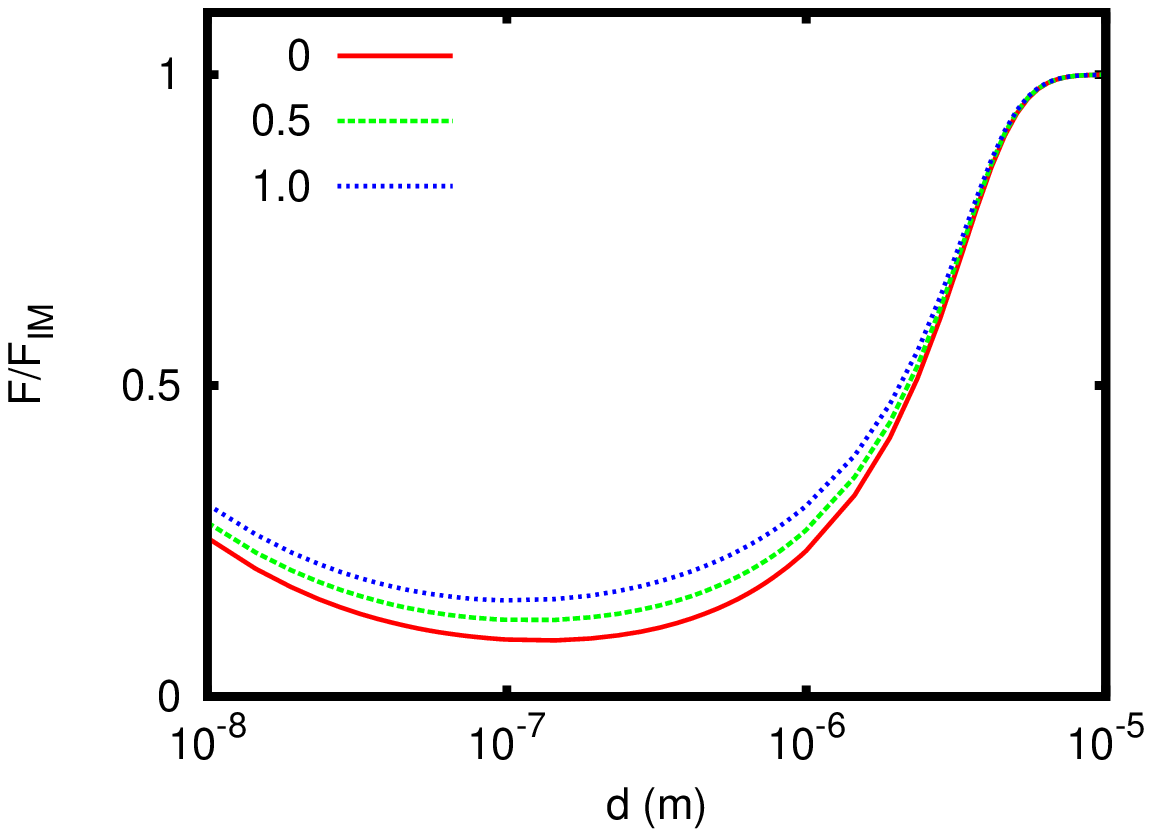, width = 0.45\textwidth}
  \caption{(Color online) Plots of the CP force of a Au nanoparticle above a suspended sheet of graphene as a function of distance $d$
           for different Fermi levels $E_F$.
           The results are normalized to the CP force of a Au nanoparticle above an ideal metal (top) or above
           a gold halfspace (bottom).\label{Fig:AuIMAu}}
\end{figure}

Now, we replace the gold halfspace by a sheet of suspended graphene. In this case (see Fig.~\ref{Fig:AuIMAu})
the force coincides with the force between the same particle and an ideal metal (ideal metal case) for large distances $d > \lambda_{\rm th}$.
This is due to the fact that for small frequencies (i.e.\ large distances) the Drude term in the conductivity 
of graphene dominates. This result has to be taken with some care, since the used model is strictly valid 
only for frequencies larger than $1/\tau$~\cite{FalkovskyVarlamov2007}. However, the frequency $1/\tau = 10^{12}\,{\rm rad/s}$ 
corresponds to a distance of about $2\,{\rm mm}$ which is much larger than the studied distances. Furthermore, this 
observation is in accordance with results found in  Ref.~\cite{KlimchitskayaMostepanenko2014} using the Dirac model 
for graphene~\cite{BordagEtAl2009}.

For small distances ($d < \lambda_{\rm th}$) the CP force is relatively 
small compared to the ideal metal case. The minimal values found are about 7\% of the ideal metal case. By 
increasing the Fermi level the force on the particle can be increased. At $d = 100\,{\rm nm}$ one can increase
the relative force from about 7\% to 12\%. Hence, the CP force between a spherical nanoparticle and a sheet of 
graphene as well as the effect of tuning is relatively small. These observations are similar to the results found for the Casimir-Lifshitz 
force between a gold halfspace and a sheet of graphene~\cite{BordagEtAl2009,Sernelius2011,Sernelius2012}. In Fig.~\ref{Fig:AuIMAu} we also show the results 
for the force between a gold nanoparticle above a sheet of graphene normalized to the case where graphene is replaced 
by a gold halfspace. The qualitative behaviour remains the same but the relative values change slightly. 

%
%
%

\section{Numerical results and discussion - spheroidal nanoparticle}

\begin{figure}[Hhbt]
  \epsfig{file = 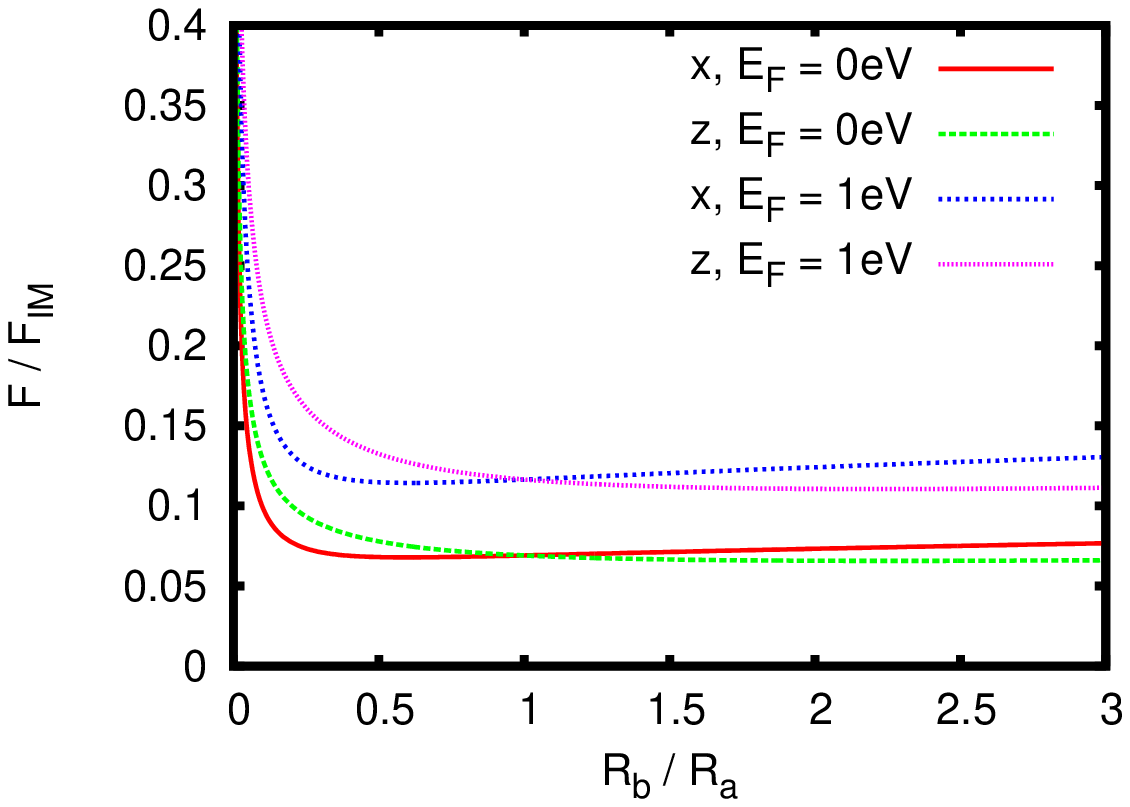, width = 0.45\textwidth}
  \epsfig{file = 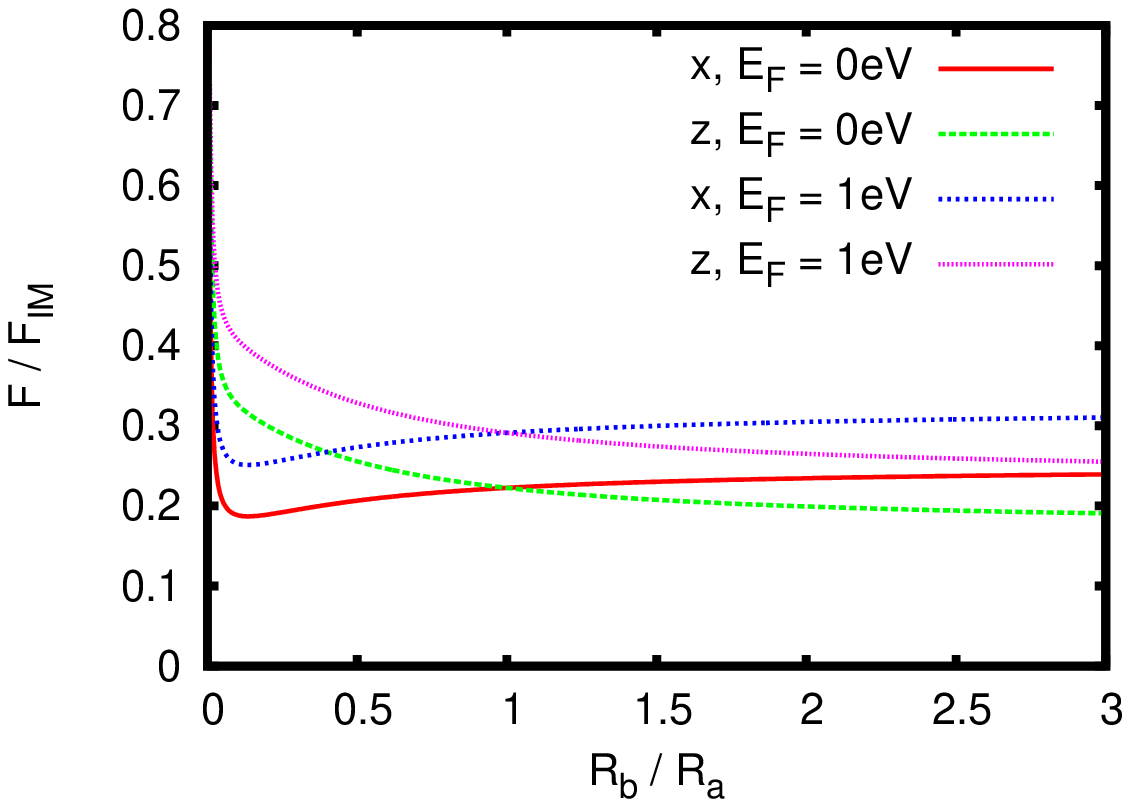, width = 0.45\textwidth}
  \caption{(Color online) Plots of the CP force of a spheroidal Au nanoparticle above a suspended sheet of graphene as a function of $R_\rb / R_\ra$
           for different Fermi levels $E_F = 0\,{\rm eV}, 1\,{\rm eV}$ and orientations (rotational axis is along $x$ or $z$ axis) choosing  
           $d = 100\,{\rm nm}$ (top) and $d = 1\,\mu m$ (bottom). The results are normalized to the CP force of the spheroidal 
           Au nanoparticle above an ideal metal.\label{Fig:SpheroidalAuIM}}
\end{figure}

To analyze the influence of the shape of the particle on the CP force we consider now a spheroidal nanoparticle
close to a suspended sheet of graphene. To start with, we study the CP force between a spheroidal gold nanoparticle and a sheet of graphene as 
function of the aspect ratios $R_\rb/R_\ra$. As can be seen in Fig.~\ref{Fig:SpheroidalAuIM} the exact CP force variation depends on the distance and the 
orientation of the nanoparticle as well as the Fermi level of the graphene sheet. For $d = 100\,{\rm nm}$ we find that with respect 
to the spherical particle the force for an oblate nanoparticle ($R_\rb/R_\ra > 1$) increases if the rotational axis is oriented parallel to the graphene 
sheet ($x$ orientation) and first decreases when the nanoparticle becomes slightly prolate ($R_\rb/R_\ra < 1$) before it 
increases rapidly for strongly prolate particles. For the nanoparticle with the rotational axis normal to the graphene
sheet ($z$ orientation) a similar trend can be seen, but the minimum in the force ratio $F/F_{\rm IM}$ is less pronounced and shifted to 
larger aspect ratios $R_\rb/R_\ra$. Similar trends for both orientations are found for $d = 1\,\mu{\rm m}$ but here
the CP force for the particle with the rotational axis normal to the surface seems to decrease monotonically with
the aspect ratio $R_\rb/R_\ra$ in the plotted region. Hence, the shape and orientation of the nanoparticle has a strong influence on the force 
excerted on that particle. Note that this effect can even be more important than the effect of tuning by changing the Fermi level 
in graphene. 

\begin{figure}[Hhbt]
  \epsfig{file = 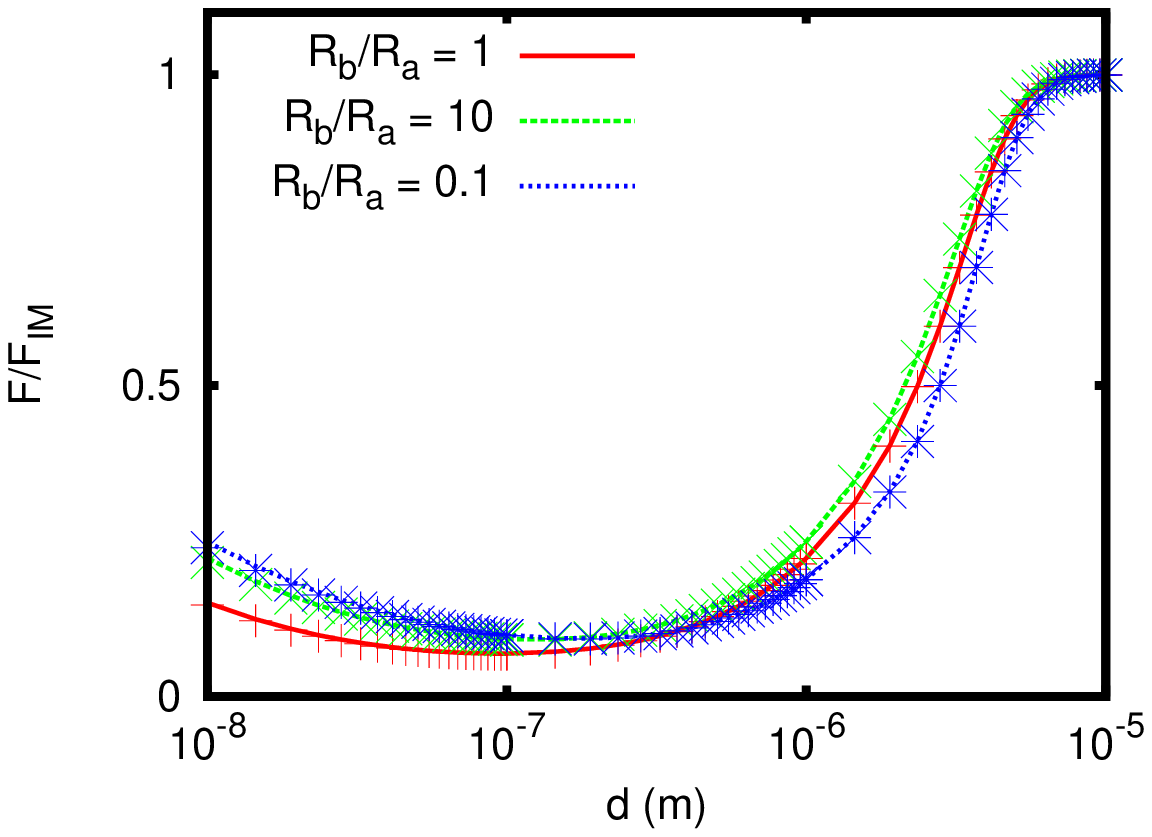, width = 0.45\textwidth}
  \epsfig{file = 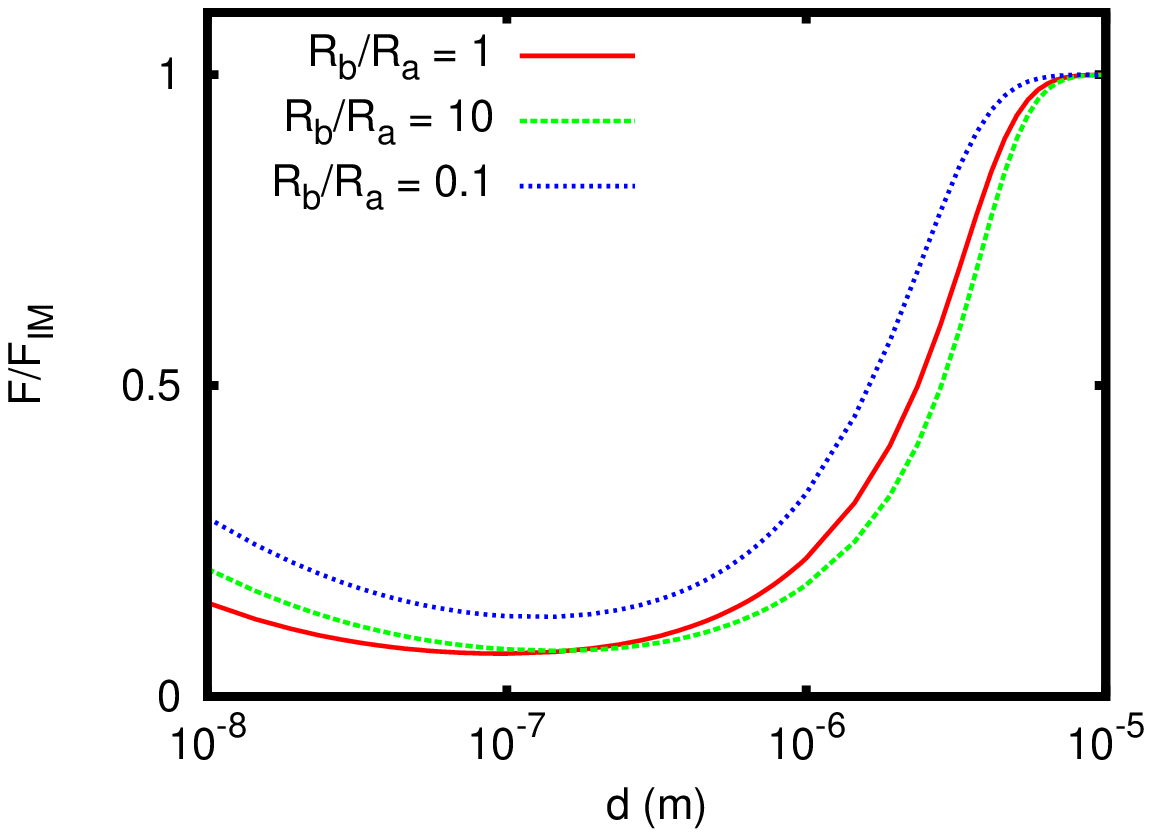, width = 0.45\textwidth}
  \caption{(Color online) Plots of the CP force of a spheroidal Au nanoparticle above a suspended sheet of graphene as a function of distance
           choosing a prolate particle with $R_\rb / R_\ra = 0.1$, an oblate particle with $R_\rb / R_\ra = 10$, and a spherical particle with $R_\rb / R_\ra = 1$.
           The Fermi level of graphene is $E_F = 0\,{\rm eV}$ and two different orientations of the nanoparticle are chosen: rotational axis is parallel
           to the graphene sheet ($x$ orientation, top) or normal to the graphene sheet ($z$ orientation, bottom). 
           The results are normalized to the CP force of the spheroidal Au nanoparticle above an ideal metal.
           Furthermore, the dots or crosses in the top figure are the results using the nonlocal reflection coefficients for graphene from Ref.~\cite{KlimchitskayaEtAl2014}.
           \label{Fig:SpheroidalAuIMDist}}
\end{figure}

In Fig.~\ref{Fig:SpheroidalAuIMDist} we show some results of the CP force between a prolate ($R_{\rm b}/R_{\rm a} = 0.1$), oblate ($R_{\rm b}/R_{\rm a} = 10$), and spherical Au nanoparticle ($R_{\rm b}/R_{\rm a} = 1$) and a sheet of graphene (here we use $E_F = 0\,{\rm eV}$) as a function of distance. It can be seen that for distances on the order of 100nm or smaller the force on the nonspherical particles is generally larger than on the spherical ones having the same volume. For larger distances the force of the spheroidal particles compared to the spherical particles is always larger when the particle axis with larger $R_a$ or $R_b$ is normal to the graphene sheet. 
On the other hand, if this particle axis with larger $R_a$ or $R_b$ is parallel to the graphene sheet, the resulting force is smaller than for a
 particle of spherical shape. As observered for sperical particles the CP force converges in all cases to the ideal metal result for $d \gg \lambda_{\rm th}$. Hence, for
such distances graphene acts like a perfect metal regardless of the shape and orientation of the nanoparticle. 

In Fig.~\ref{Fig:SpheroidalAuIMDist} we also show the results obtained by using nonlocal model for the conductivity of Graphene. 
The details of the nonlocal model are given in Ref.~\cite{KlimchitskayaEtAl2014} and are summarized in the appendix A for convenience. The nonlocal 
results are marked with dots, crosses etc.\ and show only small eviations from the local expressions for small distances which 
justifies the neglect of nonlocal effects in our work.

\begin{figure}[Hhbt]
  \epsfig{file = 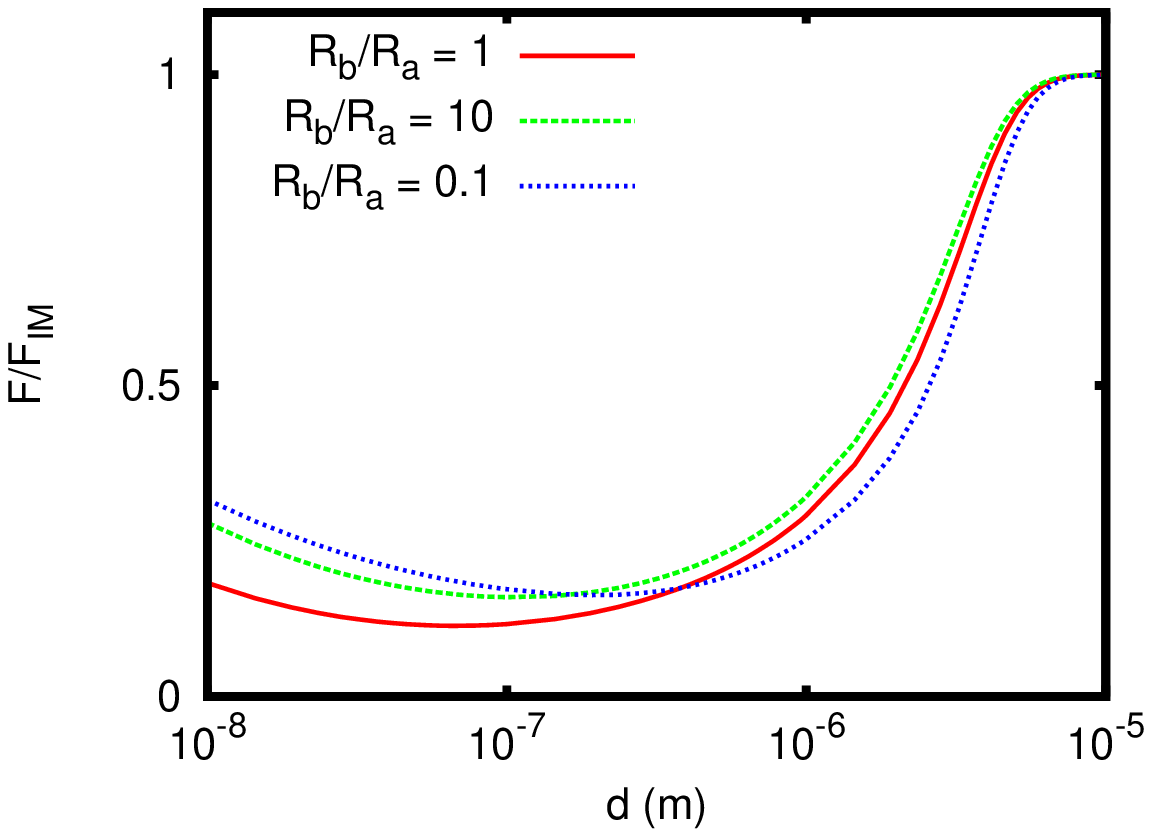, width = 0.45\textwidth}
  \epsfig{file = 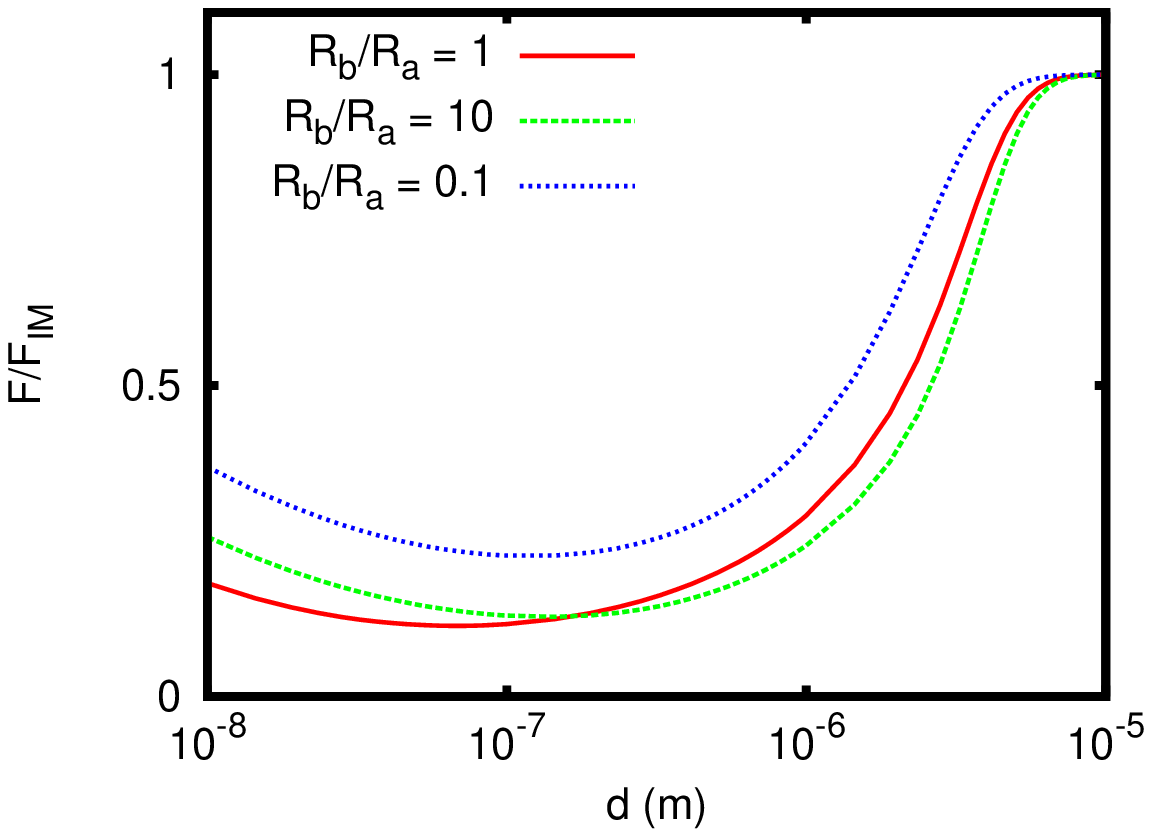, width = 0.45\textwidth}
  \caption{(Color online) As Fig.~\ref{Fig:SpheroidalAuIMDist} but with $E_{\rm F} = 1\,{\rm eV}$. \label{Fig:SpheroidalAuIMDistEF1}}
\end{figure}

In Fig.~\ref{Fig:SpheroidalAuIMDistEF1} we show the results for the same configuration as in Fig.~\ref{Fig:SpheroidalAuIMDist} but for
graphene with a Fermi level of $E_{\rm F} = 1\,{\rm eV}$. In this case, the CP force between spheroidal nanoparticles and graphene is 
much larger than between spherical particles and graphene for distances smaller than 100nm so that even values between 20-50\% of that
between a nanoparticle and a perfect metal can be achieved which is huge compared to the values of about 5\% or less found 
in Refs.~\cite{BordagEtAl2009,Sernelius2011,Sernelius2012,RibeiroScheel} and it is large compared to the value found for the spheroidal nanoparticle which 
is between 11-18\% in Fig.~\ref{Fig:SpheroidalAuIMDistEF1} of the ideal metal case.

%
%
%

\section{Conclusion}

In summary, we have considered the thermal Casimir-Polder interaction between spherical/spheroidal nanoparticles and
a sheet of graphene. We have shown that the Casimir-Polder force for spherical particles is in general small compared to the
force between the nanoparticle and a perfect metal when considering distances much smaller than the thermal wavelength. On the
other hand, for distances larger than the thermal wavelength the sheet of graphene behaves like a perfect metal a result which
was also found previously for atoms above graphene. Tuning the electron density inside the sheet of graphene by gating or
doping seems to have a relatively small impact on the resulting force. However, when considering spheroidal nanoparticles it
turns out that the shape and orientation can have an impact on the resulting force which is larger than that of the tuning inside
graphene. Depending on the distance regime the force between spheroidal nanoparticles and graphene can be larger or smaller 
than that for spherical nanoparticles having the same volume. For distances smaller then $100\,{\rm nm}$ we find that the force 
excerted on spheroidal nanoparticles is always larger than that for spherical nanoparticles with the same volume. The
effect of anisotropy and tuning together result in a force between the spheroidal nanoparticles and
graphene which can be in the range of 20-50\% of that between spheroidal nanoparticles and a perfect metal. This is quite large compared
to the Casimir force found between a gold halfspace and a sheet of graphene, between a rubidium atom and graphene, and between a spherical gold nanoparticle and 
graphene which are about 3\%, 5\% or between 11-18\% of that for perfect metals.

\appendix

\section{Nonlocal reflection coefficients for graphene}

Our nonlocal results in Fig.~\ref{Fig:SpheroidalAuIMDist} are based on the nonlocal reflection coefficients for undoped gapless 
graphene which can be expressed in terms of the polarization tensor (see for example Ref.~\cite{KlimchitskayaEtAl2014} for more details)
\begin{align}
  r_\rp &= \frac{\tilde{\gamma}_0 \Pi_{00}}{2 \hbar \kappa^2 + \tilde{\gamma}_0 \Pi_{00}}, \\
  r_\rs &= - \frac{\kappa^2 \Pi_{\rm tr} - \tilde{\gamma}_0^2  \Pi_{00}}{2 \hbar \kappa^2 \tilde{\gamma}_0 + \kappa^2 \Pi_{\rm tr} - \tilde{\gamma}_0^2  \Pi_{00}}
\end{align}
where
\begin{widetext}
\begin{equation}
\begin{split}
    \Pi_{00}(\xi_n, \kappa) &= \frac{\pi \hbar \alpha \kappa^2}{f(\xi_n,\kappa)} +\frac{8 \hbar \alpha c^2}{v_{\rm F}^2} \int_0^1\!\!\rd x\, \biggl\{ \frac{\kb T}{\hbar c} 
                            \ln \bigl[ 1 + 2 \cos(2 \pi n x) \re^{- \theta_{\rm T}(\xi_n, \kappa, x)} + \re^{-2 \theta_{\rm T}(\xi_n, \kappa, x)} \bigr] \\
                            &\quad - \frac{\xi_n}{2 c}(1 - 2x) \frac{\sin(2 \pi n x)}{\cosh\bigl(\theta_{\rm T}(\xi_n, \kappa, x)\bigr) + \cos(2 \pi n x)} 
                            + \frac{\xi_n^2}{c^2} \frac{\sqrt{x(1 - x)}}{f(\xi_n, \kappa)}  \frac{\cos(2 \pi n x) + \re^{-2 \theta_{\rm T}(\xi_n, \kappa, x)}}{\cosh\bigl(\theta_{\rm T}(\xi_n, \kappa, x)\bigr) + \cos(2 \pi n x)} 
                \biggr\}
\end{split}
\end{equation}
and
\begin{equation}
\begin{split}
    \Pi_{\rm tr}(\xi_n, \kappa) &=  \Pi_{00}(\xi_n, \kappa) + \frac{\pi \hbar \alpha}{f(\xi_n,\kappa)}\biggl[f^2(\xi_n,\kappa) + \frac{\xi_n^2}{c^2} \biggr] 
                                 + 8\hbar \alpha\int_0^1\!\!\rd x\, \biggl\{ \frac{\xi_n}{c}(1 - 2x) \frac{\sin(2 \pi n x)}{\cosh\bigl(\theta_{\rm T}(\xi_n, \kappa, x)\bigr) + \cos(2 \pi n x)} \\
                                &\quad - \frac{\xi_n^2}{c^2} \frac{\sqrt{x(1 - x)}}{f(\xi_n, \kappa)} \biggl[f^2(\xi_n,\kappa) + \frac{\xi_n^2}{c^2} \biggr] \frac{\cos(2 \pi n x) + \re^{-2 \theta_{\rm T}(\xi_n, \kappa, x)}}{\cosh\bigl(\theta_{\rm T}(\xi_n, \kappa, x)\bigr) + \cos(2 \pi n x)}. \biggr\}.
\end{split}
\end{equation}
\end{widetext}
Here, $\alpha = e^2/(4 \pi \epsilon_0 \hbar c)$ is the fine structure constant, $v_{\rm F} =  8.73723\times10^5\,{\rm m/s}$ is the Fermi velocity
in graphene and
\begin{align}
  f(\xi_n,\kappa) &= \biggl( \frac{v_{\rm F}^2}{c^2} \kappa^2 + \frac{\xi_n^2}{c^2} \biggr)^{\frac{1}{2}}, \\
  \theta_{\rm T}(\xi_n, \kappa, x) &= \frac{\hbar c}{\kb T} f(\xi_n,\kappa) \sqrt{x(1 - x)}.
\end{align}

\end{document}